\documentclass[slac_one]{revtex4}
\usepackage{graphicx}
\usepackage{fancyhdr}
\input babarsym

\begin{document}

\title{Measurement of Charmless {\boldmath $B \rightarrow VV$} decays at \babar\ } 

%
\author{Emmanuel Olaiya}
\affiliation{Rutherford Appleton Laboratory, Chilton, Didcot, OX11 OQX, UK (for the \babar\ Collaboration)}

\begin{abstract}

We present results of $B\rightarrow$ vector-vector $(VV)$ and $B\rightarrow$ vector-axial vector $(VA)$ decays $B^0 \rightarrow \phi X (X = \phi, \rho^+$ or $\rho^0)$, $B^+ \rightarrow \phi K^{(*)+}$, $B^0 \rightarrow \Kstar\Kstar$, $B^0 \rightarrow \rho^+b_1^-$ and $B^+ \rightarrow K^{*0}a_1^+$. The largest dataset used for these results is based on 465 $\times$ $10^6$  $\FourS \rightarrow$ \BB\ decays, collected with the \babar\ detector at the PEP-II $B$ meson factory located at the Stanford Linear Accelerator Center (SLAC). 
\end{abstract}

\maketitle

\thispagestyle{fancy}


\section{INTRODUCTION} 

A vector-vector or vector-axial vector $B$-meson decay can be characterised by three complex helicity amplitudes, $A_+$, $A_0$ and  $A_-$,  corresponding to meson helicity states $\lambda = +1, 0$ and $-1$, respectively. The $A_0$ amplitude is expected to dominate due to the $(V - A)$ nature of weak interactions and helicity conservation in strong interactions \cite{ali}. Most $B$ decays that arise from tree level $b \rightarrow c$ transitions follow an amplitude heirachy of $|A_0| > |A_+| > |A_-|$. The $B$ meson decays to heavy vector particles with charm such as $B \rightarrow J/\Psi K^*$ show a substantial fraction of the amplitude with transverse polarisation of the mesons ($A_\pm$). This is because the larger the meson mass, the weaker the amplitude heirachy. Therefore the amplitude heirachy $|A_0| > |A_+| > |A_-|$ is expected to be more significant in $B$ decays to light vector particles in both penguin~\cite{cheng,hchen,kagan} and tree-level~\cite{ali} transitions. In the tree-level $b \rightarrow u$ transitions, such as $B^0 \rightarrow \rho^+ \rho^-$~\cite{rhoprhom}, $B^+ \rightarrow \rho^0 \rho^+$~\cite{rho0rhom}, and  $B^+ \rightarrow \omega \rho^+$~\cite{omegarho}, the dominance of longitudinal polarisation has been confirmed by the \babar\ and Belle experiments. However a large fraction of transverse polarisation was first observed by \babar\ and confirmed by Belle for the decays $B \rightarrow \phi K^*$ and $B \rightarrow \rho K^*$, which is a significant departure from the expected dominance of the longitudinal amplitude.

This polarisation anomaly in the vector-vector decays $B \rightarrow \phi K^*$ and $B \rightarrow \rho K^*$ suggests other contributions to the amplitude, that were previously neglected. This possibility has generated great interest in  vector-vector decays, motivating a number of proposed contributions from physics beyond the standard model~\cite{beyondsm}. In addition there are mechanisms within the standard model which  have been proposed to explain the anomaly such as annihilation penguin~\cite{kagan,anhpen} or electroweak penguin, or QCD rescattering~\cite{qcdrescat}. The key to understanding this anomaly could be obtained by studying other vector-vector and vector-axial vector decays such as $B \rightarrow \phi V$, $B \rightarrow K^*K^*$, $B \rightarrow \rho b_1$, $B \rightarrow K^*a_1$ and comparing experimental results with theory.

\section{METHOD}

\begin{figure*}[t]
\centering
\includegraphics[width=95mm]{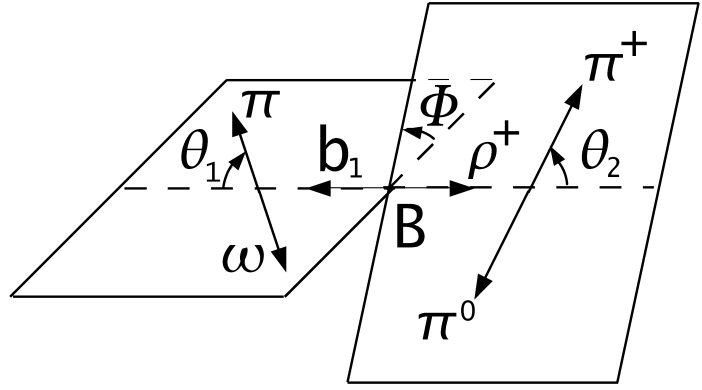}
\caption{Definition of helicity angles using the decay $B \rightarrow \rho^+ b_1$. The angle $\theta_1$ corresponds to $b_1$ and $\theta_2$ corresponds to $\rho^+$.} \label{fig:decay}
\end{figure*}

The angular distribution of the $B$ decay products can be expressed as a function of three helicity angles which describe the alignment of the particles in the decay chain. For two-body decays the polarisation is normally chosen as the direction of the daughters in the center of mass of the parent~\cite{cms}, and for three-body decays the normal to the decay plane~\cite{normal}. For $B \rightarrow VV$ and $B \rightarrow VA$ decays the angular distribution in the helicity frame is given by the equation:

\begin{equation}
\frac{d^2\Gamma }{d \cos \theta_1 \theta_2} = \frac{9}{4}[f_L \cos^2\theta_1 \cos^2\theta_2 + \frac{1}{4}( 1 - f_L) \sin^2\theta_1 \sin^2\theta_2]
\end{equation} 

where $\theta_{1,2}$ are the helicity angles and $f_L$ is the fraction of longitudinal polarisation and is equal to $|A_0|^2/(|A_0|^2 + |A_{-1}|^2 + |A_{+1}|^2)$. Here we have integrated over the azimuthal angle $\phi$ (shown in Figure~\ref{fig:decay}). We calculate the fraction of longitudinal polarisation $f_L$ by fitting  the angular distributions $\theta_1$ and $\theta_2$. 

\section{RESULTS}

\subsection{\boldmath $B^0 \rightarrow \phi\phi$, $B^+ \rightarrow \phi\rho^+$ and $B^0 \rightarrow \phi\rho^0$}
The decay $B^0 \rightarrow \phi\phi$ is an OZI suppressed process with an expected branching fraction in the range (0.1 to 3.0) $\times$ $10^{-8}$ in the standard model~\cite{btophiphi,btophiphi2,btophiphi3}. The decays $B^+ \rightarrow \phi \rho^+$ and $B^0 \rightarrow \phi\rho^0$ are pure $b \rightarrow d$ loop processes with expected branching fractions in the range (0.2 to 5.3) $\times 10^{-7}$~\cite{btophiphi4,btophiphi4a,btophiphi5,btophiphi6,btophiphi7,btophiphi8,btophiphi9,btophiphi10,btophiphi11}.


Using a data sample of 384 $\times$ $10^6$ \BB\ pairs we observed no signal for these decays, consistent with standard model expectations. We obtain a 90\% confidence level upper limit for the decays $B^0 \rightarrow \phi\phi$, $B^+ \rightarrow \phi\rho^+$ and $B^0 \rightarrow \phi\rho^0$ of 2 $\times 10^{-7}$,  30 $\times 10^{-7}$ and 3.3 $\times 10^{-7}$~\cite{bbphiv}, respectively.

\subsection{\boldmath $B^+ \rightarrow \phi K^{(*)+}$}
We search for $B^+ \rightarrow \phi K^{(*)+}_J$ decays using three final states of the $K^{(*)+}_J$ decay: $K^0_S \pi^+$, $K^+\pi^0$ and $K^+ \pi^+ \pi^-$, where $K^0_S \rightarrow \pi^+\pi^-$ and $\pi^0 \rightarrow \gamma\gamma$. Previous studies have been limited to the two body $K^*_J \rightarrow K\pi$ decays, thus considering only the spin-parity  $K^*_J$ states with P  = $(-1)^J$ . We now consider three-body states $K^*_J \rightarrow K\pi\pi$ which include  P  = $(-1)^{J+1}$ states such as $K_1$ and $K_2$. Using a data sample of 465 $\times$ $10^6$ \BB\ pairs we perform an amplitude analysis to measure branching fractions and polarisations for $B \rightarrow \phi K^*$ decays~\cite{{bbphikst}}. The results of the amplitude analysis are shown in Table~\ref{tab:phikstar}.

\begin{table}[t]
\begin{center}
\caption{Number of signal events, significance, longitudinal polarisation fraction and branching fraction measurements from the $B^{+} \rightarrow \phi K^{(*)+}$ amplitude analysis.}
\begin{tabular}{|c|c|c|c|c|}
\hline \textbf{Mode} & \textbf{$n_{sig}$} & \textbf{s ($\sigma$)} &
\textbf{$f_L$} & {\cal B} ($10^{-6}$)\\
\hline
$\phi K_1(1270)^+$ &
116 $\pm$ 26 $\pm$ 14 &
5.0&
$0.46^{+0.12 +0.03}_{-0.13 -0.07}$&
6.1 $\pm$ 1.6 $\pm$ 1.1\\
$\phi K_1(1400)^+$ &
7 $\pm$ 39 $\pm$ 17 &
0.2 &
&
$<3.2$ $(0.3 \pm 1.6 \pm 0.7$)\\
$\phi K_1(1430)^+$ &
&
5.5 &
$0.80^{+0.09}_{-0.10} \pm 0.03$&
$(8.4 \pm 1.8 \pm 0.9$)\\
~ $\rightarrow K^0_S \pi^+$&
$27 \pm 6 \pm 3$&
&
&
\\
~ $\rightarrow K^+ \pi^0$&
$39 \pm 8 \pm 4$&
&
&
\\
~ $\rightarrow K^+ \pi^+\pi^-$&
$64 \pm 14 \pm 6$&
&
&
\\
$\phi (K\pi)^{*+}_0$ &
&
8.2 &
&
$(8.3 \pm 1.4 \pm 0.8$)\\
~ $\rightarrow K^0_S \pi^+$&
$48 \pm 8 \pm 4$&
&
&
\\
~ $\rightarrow K^+ \pi^0$&
$80 \pm 13 \pm 7$&
&
&
\\
$\phi K^*_0(1430)^+$ &
&
&
&
$(7.0 \pm 1.3 \pm 0.9$)\\
$\phi K^*(1410)^+$ &
61 $\pm$ $31^{+11}_{-31}$&
$<$2.0&
&
$<$ 4.8 (2.4 $\pm$ $1.2^{+0.4}_{-1.2}$\\
$\phi K^*(1770)^+$ &
90 $\pm$ $32^{+39}_{-46}$&
$<$2.0&
&
$<$ 16.0\\
$\phi K^*(1820)^+$ &
122 $\pm$ $40^{+20}_{-83}$&
$<$2.0&
&
$<$ 23.4\\
\hline
\end{tabular}
\label{tab:phikstar}
\end{center}
\end{table}

\subsection{\boldmath $B^0 \rightarrow \Kstarz\Kstarzb$ and $B^0 \rightarrow \Kstarp\Kstarm$}


Decays proceeding via electroweak and gluonic $b \rightarrow d$ penguin diagrams have been measured in the decays $B \rightarrow \rho\gamma$~\cite{rhogamma} and $B^0 \rightarrow \Kz \Kzb$~\cite{kzkzb}. However, the decay $B^0 \rightarrow \Kstarz\Kstarzb$  proceeds via both electroweak and gluonic $b \rightarrow d$ penguin loops to two vector particles. Theoretical calculations for $B^0 \rightarrow \Kstarz\Kstarzb$ branching fractions cover a range (0.16-0.96) $\times 10^{-6}$~\cite{btophiphi6,btophiphi8,kszkszb}. The latest predictions by Beneke {\em et al.}~\cite{btophiphi3} estimate a branching fraction of $(0.6^{+0.1+0.3}_{-0.1-0.2}) \times 10^{-6}$ and $f_L$ = 0.69 $\pm$ $0.01^{+0.16}_{-0.20}$. Using a data sample of 383 $\times$ $10^6$ \BB\ pairs we observe the decay $B^0 \rightarrow \Kstarz\Kstarzb$ for the first time, measuring a branching fraction of  $(1.28^{+0.35}_{-0.30} \pm 0.11) \times 10^{-6}$ and  $f_L$ = $0.80^{+0.10}_{-0.12} \pm 0.06$~\cite{bbkzkz}, which is consistent with theoretical predictions.

The decay $B^0 \rightarrow \Kstarp\Kstarm$ is expected to occur through a $b \rightarrow u$ quark transition via a $W$-exchange, and consequently have a small branching fraction with predictions of  $(0.09^{+0.05+0.12}_{-0.03-0.10}) \times 10^{-6}$~\cite{btophiphi3} and $(0.1 \pm 0.0 \pm 0.1) \times 10^{-6}$~\cite{cheng} both based on QCD factorisation. Using a data sample of 454 $\times$ $10^6$ \BB\ pairs we observe no signal and obtain a 90\% confidence upper limit of 2.0 $\times 10^{-6}$~\cite{bbkpkm} which is two orders of magnitude more sensitive than the previous measurement~\cite{kpkml} .

\subsection{\boldmath $B^0 \rightarrow \rho^+b_1^-$}

Cheng and Yang~\cite{cheng} have extended calculations by Beneke {\em et al.}\cite{btophiphi3} to include vector-axial vector ($VA$) decays. They predict a $B^0 \rightarrow \rho^+b_1^-$ branching fraction of  $(32.7^{+16.5+12.1}_{-14.7-4.7}) \times 10^{-6}$. This measurement combines the branching fractions of both $B^0 \rightarrow \rho^+b_1^-$ and $B^0 \rightarrow \rho^-b_1^+$, noting that the latter decay is suppressed by $G$-parity conservation. They also predict the longitudinal polarisation fraction to be $f_L$ = $0.96^{+0.01}_{-0.02}$.

Using a data sample of 465 $\times$ $10^6$ \BB\ pairs we find no evidence of a signal and obtain a 90\% confidence level upper limit of 1.7 $\times 10^{-6}$~\cite{bbb1rho}.

\subsection{\boldmath $B^+ \rightarrow K^{*0}a_1^+$}

Recent searches of charmless decays with axial-vectors have revealed comparatively large branching fractions compared to other charmless decays, in the range $(15-35) \times 10^{-6}$~\cite{btova}. Available $B^+ \rightarrow K^{*0}a_1^+$ theoretical estimates are based on na\"{i}ve factorisation with an expected branching fraction of 0.51 $\times 10^{-6}$~\cite{calderon} and on QCD factorisation with an expected branching fraction of $(9.7^{+4.9+32.9}_{-3.5-2.4}) \times 10^{-6}$ and a longitudinal polarisation fraction $f_L$ of $0.38^{+0.51}_{-0.40}$~\cite{cheng}.

Using a data sample of 465 $\times$ $10^6$ \BB\ pairs we find no evidence for signal and measure a branching fraction of $(1.4^{+0.8+1.4}_{-1.0-1.4}) \times 10^{-6}$ corresponding to a 90\% confidence level upper limit of 3.2 $\times 10^{-6}$~\cite{bba1kst}.

\section{CONCLUSION}

Using larger datasets, the \babar\ experiment has provided more precise $B \rightarrow VV$ measurements, further supporting the smaller than expected longitudinal polarisation fraction of $B \rightarrow \phi K^*$. Additional $B$ meson to vector-vector and vector-axial vector decays have also been studied with a view to shedding light on the polarisation anomaly. Taking into account the available errors, we find no disagreement between theory and experiment for these additional decays.


\begin{acknowledgments}

I would like to thank the organisers of the ICHEP 08 conference, the \babar\ collaboration and the UK Science and Technology Facilities Council (STFC).

\end{acknowledgments}



\begin{thebibliography}{99} 

\bibitem{ali}
A. Ali {\em et al.}, Z. Physik {\bf C1}, 269 (1979);\\
M. Suzuki, Phys. Rev. {\bf D64}, 117503 (2001).

\bibitem{cheng}
H. Y. Cheng and K. C. Yang, Phys. Lett. {\bf B511}, 40 (2001).

\bibitem{hchen}
C. H. Chen, Y. Y. Keum and H. n. Li, Phys. Rev. {\bf D66}, 054013 (2002).

\bibitem{kagan}
A. L. Kagan, Phys. Lett {\bf 601}, 151 (2004);\\
Y. Grossman, Int. J. Mod. Phys. {\bf A19}, 907 (2004).

\bibitem{rhoprhom}
Belle Collaboration, A. Somov {\em et al.}, Phys. Rev. Lett. {\bf 96}, 171801 (2006);\\ \babar\ Collaboration, B. Aubert {\em et al.}, Phys. Rev. {\bf D76}, 052007 (2007).

\bibitem{rho0rhom}
Belle Collaboration, J. Zhang {\em et al.}, Phys. Rev. Lett. {\bf 91}, 221801 (2003);\\ \babar\ Collaboration, B. Aubert {\em et al.}, Phys. Rev. Lett {\bf 97}, 261801 (2006).

\bibitem{omegarho}
\babar\ Collaboration, B. Aubert {\em et al.}, Phys. Rev. {\bf D74}, 051102 (2006).

\bibitem{beyondsm}
E. Alvarez {\em et al.}, Phys. Rev. D {\bf 70}, 115014 (2004); P. K. Das and K. C. Yang, Phys. Rev. D {\bf 71}, 094002 (2005); C. H. Chen and C. Q. Geng, Phys. Rev. D {\bf 71}, 115004 (2005).

\bibitem{anhpen}
H. n. Li and S. Mishima, Phys. Rev. D {\bf 71}, 054025 (2005); C.-H. Chen {\em et al.}, Phys. Rev. D {\bf 72}, 054011 (2005); M. Beneke {\em et al.}, Phys. Rev. Lett. {\bf 96}. 141801 (2006), arXiv:hep-ph/0612290; C.-H. Chen and C.-Q Geng, Phys. Rev. D 75, 054010 (2007); A. Datta {\em et al.}, arXiv:0705.3915[hep-ph]. 

\bibitem{qcdrescat}
C. W. Bauer {\em et al.}, Phys Rev. D {\bf 70}, 054015 (2004); P. Colangelo {\em et al.}, Phys. Lett. B {\bf 597}, 291 (2004).

\bibitem{cms}
M. Jacob and G. C. Wick, Ann. Phys. {\bf 7}, 404 (1959).

\bibitem{normal}
S. M. Berman and M. Jacob, Phys. Rev. {\bf 139}, 1023 (1965).

\bibitem{btophiphi}
S. Bar-Shalom, G. Eilam, and Y. D. Yang, Phys. Rev. D {\bf 67}, 014007 (2003).

\bibitem{btophiphi2}
C. D. Lu {\em et al.}, Eur. Phys. Jour. C {\bf 41}, 311 (2005).

\bibitem{btophiphi3}
M. Beneke, J. Rohrer and D. Yang, Nucl. Phys. B {\bf 774}, 64 (2007).

\bibitem{btophiphi4}
D. Du and Z. Xing, Phys. Lett. B {\bf 312}, 199 (1993).

\bibitem{btophiphi4a}
M. Gronau and J. L. Rosner, arXiv:0806.3584.

\bibitem{btophiphi5}
A. Deandrea {\em et al.}, Phys. Lett. B {\bf 320}, 170 (1994).

\bibitem{btophiphi6}
A. Ali {\em et al.}, Phys. Rev D {\bf 58}, 094009 (1998).

\bibitem{btophiphi7}
Y. H. Chen {\em et al.}, Phys. Rev. D {\bf 60}, 094014 (1999).

\bibitem{btophiphi8}
W. Zou and Z. Xiao, Phys. Rev. D{\bf 72}, 094026 (2005).

\bibitem{btophiphi9}
C. D. Lu {\em et al.}, Chin. Phys. Lett. {\bf 23}, 2684 (2006).

\bibitem{btophiphi10}
J. Li. {\em et al.} hep-ph/0607249.

\bibitem{btophiphi11}
S. Bao, F. S. Su, Y.-L Wu, and C.Zhuang, arXiv:0801.2596.

\bibitem{bbphiv}
B. Aubert {\em et al.} (\babar\ Collaboration), arXiv:0807.3935 [accepted for publication in PRL].

\bibitem{bbphikst}
B. Aubert {\em et al.} (\babar\ Collaboration), arXiv:0806.4419  [accepted for publication in PRL].

\bibitem{rhogamma}
B. Aubert {\em et al.} (\babar\ Collaboration), Phys. Rev. Lett. {\bf 98}, 151802 (2007); D. Mohapatra {\em et al.} (Belle Collaboration), Phys. Rev. Lett. {\bf 96}, 221601 (2006).

\bibitem{kzkzb}
B. Aubert {\em et al.} (\babar\ Collaboration), Phys. Rev. Lett. {\bf 97}, 171805 (2006); S.-W. Lin {\em et al.} (Belle Collaboration) Phys. Rev. Lett.  {\bf 98}, 181804 (2007).

\bibitem{kszkszb}
L. Chau {\em et al.}, Phys. Rev. D {\bf 45}, 3143 (1992).

\bibitem{bbkzkz}
B. Aubert {\em et al.} (\babar\ Collaboration), Phys. Rev. Lett. {\bf 100}, 081801 (2008).

\bibitem{bbkpkm}
B. Aubert {\em et al.} (\babar\ Collaboration), Phys. Rev. D {\bf 78}, 051103 (2008).

\bibitem{kpkml}
R. Godang {\em et al.} (CLEO Collaboration), Phys. Rev. Lett. {\bf 88}, 021802 (2002).

\bibitem{bbb1rho}
W.T.Ford, ``Hadronic $B_u$ and $B_d$ decays'', arXiv:0810.0494.

\bibitem{btova}
 \babar\ Collaboration, B. Aubert {\em et al.}, Phys. Rev. Lett {\bf 99}, 261801 (2007); B. Aubert {\em et al.}, Phys. Rev. Lett {\bf 100}, 051803 (2008).

\bibitem{calderon}
G. Calder\'{o}n, J. H. Mun\~{o}z and C. E. Vera, Phys. Rev. D {\bf 76}, 094019 (2007).

\bibitem{bba1kst}
B. Aubert {\em et al.} (\babar\ Collaboration), arXiv:0808.0579.
\end{thebibliography}
\end{document}